\documentclass[twocolumn]{aastex631}

\shorttitle{Symbiotic Stars are Not Supernova Progenitors}
\shortauthors{Schaefer}
\graphicspath{{./}{figures/}}

\begin{document}
\title{Symbiotic Stars (Including T Corona Borealis) Are Not Immediate Progenitors of Normal Type {\rm I}a Supernovae}

\author[0000-0002-2659-8763]{Bradley E. Schaefer}
\affiliation{Department of Physics and Astronomy,
Louisiana State University,
Baton Rouge, LA 70803, USA}

\begin{abstract}

A popular solution to the Type Ia supernova (SNIa) progenitor problem is that the immediate progenitors are symbiotic star systems.  This solution requires that the companion star of the exploding white dwarf must be a red giant star with a heavy stellar wind.  This has been tested for 189 normal SNIa, with {\it all} tested systems being proven to not have the required red giant:  (A) Zero-out-of-9 normal type Ia supernova remnants have any red giant ex-companion star near the center with limits of $M_V$$>$0.0.  (B) Zero-out-of-2 normal SNIa in nearby galaxies have any red giant at the position as seen in archival pre-eruption images by {\it HST} to limits of $M_V$$>$0.0.  (C and D) Zero-out-of-111 normal SNIa have any detected hydrogen or helium emission lines in their eruption spectra, with limits on entrained gas of $M_{\rm H}$$<$0.22 and $M_{\rm He}$$<$0.07 M$_{\odot}$, which is the minimum mass lost by a red giant in a nearby blastwave.  (E and F) Zero-out-of-9 nearby normal SNIa were detected in the radio or X-rays, as required from the ejecta/wind impact, to limits of $\dot{M}_{\rm wind}$$<$3$\times$$10^{-9}$ M$_{\odot}$ yr$^{-1}$.  (G) Zero-out-of-$\sim$69 normal SNIa display any brightening in the first few days to limits of $M_V$$>$$-$18, as required for a red giant companion when we are looking down its shadowcone.  With zero-out-of-189 normal SNIa having any possibility of having a red giant companion, the fraction of SNIa with symbiotic progenitors is $<$0.53\%.  The overwhelming conclusion is that normal SNIa are not from symbiotic-progenitors in any measurable fraction.

\end{abstract}

\section{INTRODUCTION}

The nature of the immediate progenitor systems of Type Ia supernovae (SNIa) is a grand challenge problem of high importance for all of astrophysics, where for decades there has been vigorous experiments, theory, and discussion, yet with no final conclusion (Livio 2000, Maoz, Mannucci, \& Nelemans 2014, Livio \& Mazzali 2018, Patat \& Hallakoun 2018, Ruiter \& Seitenzahl 2025).  Everyone agrees that the SNIa progenitors involve a close binary system with a white dwarf (WD) of carbon/oxygen (CO) composition.  But there are many possibilities for the nature of the companion star close around the WD, how the mass of the CO WD increases to the point of igniting runaway thermonuclear burning so as to explode the WD, and the detailed site and explosion mechanism within the WD.  All popular solutions to the SNIa-progenitor problem fall into one of two groups of explanations, with these groups being labeled by the number of WDs in the binary.  The single-degenerate (SD) models all postulate that there is only one WD (i.e., only one degenerate star in the binary), where the other star in the binary is a relatively normal companion star that supplies gas, usually by Roche lobe overflow, to the CO WD, leading to its ultimate explosion.  The double-degenerate (DD) models postulate that the companion star is a second WD (i.e., with two degenerate stars) that is in-spiraling (from gravitational radiation) so that the two WDs collide or merge, whereupon the carbon is ignited to power the complete disintegration of the star.

One of the most popular SD models is that the progenitor is a symbiotic star, where the companion is a red giant star with a massive stellar wind feeding the WD (Kenyon et al. 1993, Hachisu, Kato, \& Nomoto 1999, Meng \& Yang 2010).  Symbiotic star systems (SSS) consist of a WD in a fairly close binary with a red giant that has a dense stellar wind (Miko{\l}ajewska 2010, Munari 2019, Miko{\l}ajewska et al. 2021).  One class of SSS is the S-type, where the red giant fills its Roche lobe with this supplying the gas that falls onto the WD.  The other class of SSS is the D-type, where a high density wind provides the gas that falls onto the WD.  A fraction of the SSS have companions that are ordinary Mira stars, with these all being D-type.  The SSS appear in the optical as dominated by the spectrum of the red giant, but with a blue excess from the accretion light, with the usual flickering in the light curves on all time scales.  The stellar wind and its dust makes the SSS bright in the infrared and radio bands.  The accretion onto the WD makes for the SSS to be weak X-ray sources.  T Corona Borealis (T CrB) is the most famous SSS, with an M4 III companion in a 227 day orbit, and with known classical nova eruptions in the years 1217, 1787, 1866, 1946, and one that is likely to erupt any month now. (Schaefer 2023a, 2023b).  T CrB is just one of four so-called `symbiotic recurrent nova' (SyRN) systems of four similar sister systems, with the others being RS Oph, V3890 Sgr, and V745 Sco.

The SSS-progenitor model has the requirement that the supernova progenitors have an ordinary red giant companion with a heavy stellar wind.  (This is just saying that the progenitor is an SSS.)  This requirements can be tested by seeking for a red giant companion star by all of seven search methods (Section 2).  The two direct tests seek to spot the red giant star either {\it before} or {\rm after} the supernova (Sections 2.1 and 2.2).  The five indirect tests seek to spot some inevitable effect caused by any red giant caught in the blastwave from the exploding WD.  These effects include bright nebular {\it hydrogen} and {\it helium} lines from the red giant atmosphere stripped and entrained into the eject (Section 2.3), the impact of the SN ejecta onto the stellar wind around the red giant making highly luminous {\it radio} and {\it X-ray} sources (Section 2.4), and the Kasen effect (Kasen 2010) where a red giant creates a hole in the ejecta whose edge is bright shocked gas creating a blatant peak in the light curve with the peak reaching comparable to the main maximum within hours of the explosion (Section 2.5).

\section{TESTING WHETHER SN{\rm I}\lowercase{a} HAVE A RED GIANT COMPANION STAR}

The SSS-progenitor model requires that the companion star be a red giant in a fairly close orbit around the WD.  This makes a prediction that can be tested.  If a particular SNIa is found to have a red giant companion, then this would substantially strengthen the SSS-progenitor model.  If a particular SNIa is found to not have a red giant companion, then that progenitor cannot be a SSS.  If many SNIa are found to have no red giant companions, then the fraction of SSS-progenitors must be small.

A number of methods have been used to test whether a particular SNIa  has a red giant companion star.  Here, I will enumerate the many methods and particular supernovae tested, with the results summarized in Table 1.  Note that these tests are all made for normal SNIa events, with no rare or unusual event classes.  (That is, the tests here do not include any of the various types of non-normal SNIa, including Iax, Ia-CSM, or SLSN-I eruptions, see Liu et al. 2023.  These classes are rare and uncommon as measured by the fraction seen in the sky, and are irrelevant to the general SNIa progenitor problem and irrelevant to the supernova cosmology research.  These not-normal events are by some completely different mechanism and evolution from normal SNIa events.  For this paper, the only important issue is to be sure that the data are only taken for normal SNIa events.)  By having the testing confined to normal SNIa, this is addressing the progenitor problem specifically.

\begin{table*}
	\centering
	\caption{Do SN{\rm I}a Progenitors have a Giant Companion?}
	\begin{tabular}{llccll}
		\hline
		Method   &   Supernova  &  \#tested   &   \#RG$^a$   &   Constraint   &  Ref.$^b$ \\
		\hline
Ex-companion	&	Tycho's SN	&	1	&	0	&	$M_V$ $\ge$ $+$2.4	&	[1]	\\
Ex-companion	&	Kepler's SN	&	1	&	0	&	$M_R >$ $+$3.4	&	[2]	\\
Ex-companion	&	SN 1006	&	1	&	0	&	$M_V >$ $+$4.9	&	[3]	\\
Ex-companion	&	SNR G272.2-3.2	&	1	&	0	&	$M_G >$ $+$7.6	&	[4]	\\
Ex-companion	&	SNR 0509-67.5 in LMC	&	1	&	0	&	$M_V >$ $+$8.4	&	[5]	\\
Ex-companion	&	SNR 0519-69.0 in LMC	&	1	&	0	&	$M_V >$ $+$1.2	&	[6]	\\
Ex-companion	&	SNR 0505-67.9 in LMC	&	1	&	0	&	$M_V >$ $+$0.6	&	[7]	\\
Ex-companion	&	SNR 0509-68.7 in LMC	&	1	&	0	&	$M_V >$   ~~0.0	&	[7]	\\
Ex-companion	&	SN 1972E in NGC 5253	&	1	&	0	&	$M_V >$ $+$0.5	&	[8]	\\
Pre-eruption progenitor	&	SN 2011fe in M101	&	1	&	0	&	$M_V >$ $+$1.2	&	[9]	\\
Pre-eruption progenitor	&	SN 2014J in M82	&	1	&	0	&	$M_V >$ $-$1	&	[10]	\\
Nebular H and He lines	&	SN 2017cbv in NGC 5643	&	1	&	0	&	$M_{H}$$<$0.0001, $M_{He}$$<$0.0005 M$_{\odot}$	&	[11]	\\
Nebular H and He lines	&	SN 2018oh in UGC 4780	&	1	&	0	&	$M_{H}$$<$0.00054, $M_{He}$$<$0.000047 M$_{\odot}$	&	[12]	\\
Nebular H and He lines	&	SN 2020nlb in M85	&	1	&	0	&	$M_{H}$$<$0.002, $M_{He}$$<$0.004 M$_{\odot}$	&	[13]	\\
Nebular H and He lines	&	SN 2021aefx in NGC 1566	&	1	&	0	&	$M_{H}$$<$0.0006, $M_{He}$$<$0.005 M$_{\odot}$	&	[14]	\\
Nebular H lines	&	107 normal SN{\rm I}a	&	107	&	0	&	Entrained $M_{H}$$<$0.22 M$_{\odot}$	&	[15]	\\
Nebular He lines	&	93 normal SN{\rm I}a	&	93	&	0	&	Entrained $M_{He}$$<$0.07 M$_{\odot}$	&	[15]	\\
Ejecta/Wind impact in radio	&	SN 2011by in NGC 3972	&	1	&	0	&	$\dot{M}_{\rm wind} <$ 3$\times$$10^{-9}$ M$_{\odot}$ yr$^{-1}$	&	[16]	\\
Ejecta/Wind impact in radio	&	SN 2011fe in M101	&	1	&	0	&	$\dot{M}_{\rm wind} <$ 2.4$\times$$10^{-10}$ M$_{\odot}$ yr$^{-1}$	&	[16]	\\
Ejecta/Wind impact in radio	&	SN 2012cg in NGC 4424	&	1	&	0	&	$\dot{M}_{\rm wind} <$ 1.5$\times$$10^{-9}$ M$_{\odot}$ yr$^{-1}$	&	[16]	\\
Ejecta/Wind impact in radio	&	SN 2012fr in NGC 1365	&	1	&	0	&	$\dot{M}_{\rm wind} <$ 2.4$\times$$10^{-9}$ M$_{\odot}$ yr$^{-1}$	&	[16]	\\
Ejecta/Wind impact in radio	&	SN 2012ht in NGC 3447	&	1	&	0	&	$\dot{M}_{\rm wind} <$ 2.4$\times$$10^{-9}$ M$_{\odot}$ yr$^{-1}$	&	[16]	\\
Ejecta/Wind impact in radio	&	SN 2014J in M82	&	1	&	0	&	$\dot{M}_{\rm wind} <$ 3$\times$$10^{-10}$ M$_{\odot}$ yr$^{-1}$	&	[16]	\\
Ejecta/Wind impact in radio	&	SN 2021aefx in NGC 1566	&	1	&	0	&	$\dot{M}_{\rm wind} <$ 3$\times$$10^{-9}$ M$_{\odot}$ yr$^{-1}$	&	[14]	\\
Ejecta/Wind impact in X-rays	&	SN 2017cbv in NGC 5643	&	1	&	0	&	$\dot{M}_{\rm wind} <$ 2.2$\times$$10^{-9}$ M$_{\odot}$ yr$^{-1}$	&	[11]	\\
Ejecta/Wind impact in X-rays	&	SN 2020nlb in M85	&	1	&	0	&	$\dot{M}_{\rm wind} <$ 2.9$\times$$10^{-9}$ M$_{\odot}$ yr$^{-1}$	&	[11]	\\
Kasen effect	&	307 normal SN{\rm I}a	&	$\sim$31	&	0	&	{\it TESS} light curves	&	[17]	\\
Kasen effect	&	122 normal SN{\rm I}a	&	$\sim$12	&	0	&	ZTF light curves	&	[18]	\\
Kasen effect	&	108 normal SN{\rm I}a	&	$\sim$11	&	0	&	SDSS--{\rm II} light curves	&	[19]	\\
Kasen effect	&	87 normal SN{\rm I}a	&	$\sim$9	&	0	&	SNLS light curves	&	[20]	\\
Kasen effect	&	61 normal SN{\rm I}a	&	$\sim$6	&	0	&	LOSS light curves	&	[21]	\\
Kasen effect	&	4 normal SN{\rm I}a	&	$\sim$0.4	&	0	&	{\it Kepler} light curves	&	[22], [23]	\\
		\hline
	\end{tabular}	
	
\begin{flushleft}	
\
$^a$ Number out of the tested normal SNIa for which a red giant companion is possible.\\
$^b$ References: [1] Xue \& Schaefer (2015), 
[2] Ruiz-Lapuente et al. (2018), 
[3] Gonz\'{a}lez Hern\'{a}ndez et al. (2012), 
[4] Ruiz-Lapuente et al. (2023), 
[5] Schaefer \& Pagnotta (2012), 
[6] Edwards, Pagnotta, \& Schaefer (2012), 
[7] Pagnotta \& Schaefer (2015), 
[8] Do et al. (2021), 
[9] Tucker \& Shappee (2024), 
[10] Kelly et al. (2014), 
[11] Sand et al. (2021), 
[12] Dimitriadis et al. (2019a), 
[14] Hosseinzadeh et al. (2022), 
[15] Tucker et al. (2020), 
[16] Chomiuk et al. (2016), 
[17] Fausnaugh et al. (2023), 
[18] Yao et al. (2019), 
[19] Hayden et al. (2010), 
[20] Bianco et al. (2011), 
[21] Ganeshalingam et al. (2011), 
[22] Olling et al. (2015), 
[23] Wang et al. (2021). 
\\
 
\end{flushleft}	
\end{table*}

\subsection{No Red Giant Ex-companion Stars}

If an SSS explodes as a supernova, the ex-companion star will remain close to the explosion site for many centuries.  This red giant will be battered by the nearby supernova, with a substantial fraction of its atmosphere being ablated by the blastwave (Livne, Tuchman, \& Wheeler 1992, Marietta, Burrows, \& Fryxell 2000, Podsiadlowski 2003, Pan, Ricker, \& Taam 2012, Chen, Meng, \& Han 2017, Wang, Chen, \& Pan 2025).  The remaining atmosphere will expand and heat up, with this extended hydrogen-rich envelope appearing comparable to the original red giant.  The core of the star will be shock-heated, so its luminosity will be increased immediately after the blast.  In the long term, the core of the ex-companion is largely unaffected and continues to shine near its pre-supernova luminosity.  Detailed calculations show that the luminosity and surface temperature have relatively small transient changes, lasting for many centuries until the companion settles to a case near its pre-supernova state.  Depending on the exact circumstances, for the centuries after the supernova, these calculations show that red giant ex-companions will typically brighten-or-dim by one or two magnitudes, with the surface temperature remaining relatively unchanged.  So the ex-companion star of an SSS progenitor will appear much like a red giant star.

After the supernova ejecta clears, any SSS ex-companion should appear as a red giant near the center of the expanding supernova remnant.  This is testable.  For a useful test, the supernova remnant must be provable from a normal Type Ia supernova (for example, by the spectra of light echoes, see Rest et al. 2005, 2008a, 2008b), the remnant must be relatively young (so that the ex-companion does not have time to get far from the explosion site), and the remnant must be moderately close (so that the ex-companion can be recognized well above detection limits).  Few supernova remnant fulfill these test conditions (Ruiz-Lapuente 2019).  Table 1 lists the results for nine such tests.  In all cases, no ex-companion has been identified to deep limits.

The first limit, and still by far the deepest, comes from SNR 0509-67.5 in the LMC.  This is a young remnant with a simple ring structure, so the explosion site can be accurately plotted as the center of the remnant.  Rest et al. (2005, 2008a) made a wonderful discovery of a light echo from the original supernova, proving that the supernova spectrum was of a normal Type Ia supernova, plus giving an age of 400$\pm$50 years.  Being in the LMC, the remnant is in an uncrowded field with low extinction.  A red giant ex-companion star would appear at near 18th magnitude.  SNR 0509-67.5 is the optimal case for detecting any ex-companion.  Schaefer \& Pagnotta (2012) used deep images from {\it HST} to search for any star within 1.43 arc-seconds of the remnant center, with this being the extreme 3$\sigma$ edge of the error circle for maximal velocity of the ex-companion.  This error circle must contain any ex-companion star, but the {\it HST} images show the circle to be empty of all stars\footnote{There is a very faint $z$=0.031 galaxy in the error circle (Pagnotta, Walker, \& Schaefer (2014).} to a limit of $V$$>$26.9.  This is a stark limit on any ex-companion, with the absolute magnitude required to be fainter than $M_V$ $>$ $+$8.4.  There is no way to get out of this severe limit.  This limit rejects any giant companions, any sub-giant companions, and any main sequence stars down to M0.  For this one supernova, this limit rejects nearly all possible single-degenerate models.  For purposes of this paper, this limit certainly rules out the possibility of a SSS-progenitor.

The idea to search for ex-companions started with Ruiz-Lapuente (1997), with the first application to Tycho's supernova of 1572 and its bright remnant (Ruiz-Lapuente et al. 2004).  Detailed analysis of the light curve, radio emission, and X-ray spectrum are convincing that the supernova was a normal Type Ia event (Ruiz-Lapuente 2004).  Ruiz-Lapuente et al. (2004) searched optical images from the ground and from {\it HST} within 39 arc-seconds of the remnant center (as measured from the {\it Chandra} image).  They identified a sub-giant star (labeled star `G') in the correct distance range that had a high space velocity, and they concluded that this star was the ex-companion from a progenitor much like the recurrent nova U Sco.  This claim excited a series of counter-arguments and replies going several levels deep (e.g., Kerzendorf et al. 2009, Gonz\'{a}lez Hern\'{a}ndez et al. 2009, Kerzendorf et al. 2013, Bedin et al. 2014).  But the identity of star G as the ex-companion failed because it turned out to be far from the astrometric position of the supernova and to be far from the expansion center of the remnant (Xue \& Schaefer 2015).  Of the stars anywhere near the 3$\sigma$ position of the explosion site (stars O, P, F, and N in Bedin et al. 2014), the {\it Gaia} DR3 distances are 0.83 kpc for star O, 4.61 kpc for star P, 1.86 kpc for star F, and 4.42 kpc for star N.  The distance to the supernova of 1572 is 2.35$\pm$0.20 kpc from Schaefer (1996) and 2.83$\pm$0.79 kpc from Ruiz-Lapuente (2004).  Based on distance, stars P, F, and N cannot be rejected at the 3$\sigma$ confidence level if we choose the supernova distance with the largest uncertainty.  These candidates do not have high space velocity, so they are poor possibilities.  For these marginally acceptable candidates, with the {\it Gaia} distances and $E(B-V)$=0.60 mag, the $M_V$ values are $+$2.4 for star P, $+$4.5 for star F, and $+$3.2 for star N.  So for the brightest possible candidates, the limit for ex-companions is $M_V$ $\ge$ $+$2.4.  With this, any red giant ex-companion is rejected.  Another way of seeing the same result is to calculate that a red giant (with $M_V$$<$0.0) at the distance of the supernova should appear at $V$$<$13.8, and there are no such stars in the area.  The progenitor of Tycho's supernova was not an SSS.

Kepler's supernova of 1604 provides a nice opportunity to spot any ex-companion star, due to it having a young and nearby remnant.  The nature of the supernova was uncertain for a long time, but this was resolved by Reynolds et al. (2007) when they measured the characteristic O/Fe ratio in the X-ray spectrum of the remnant.  Kerzendorf et al. (2014) and Ruiz-Lapuente et al. (2018) have both examined deeply in the central regions of the Kepler supernova remnant for any ex-companion star.  They show that any ex-companion must be less luminous than $M_R$ $>$ $+$3.4.  For their conclusions, ``We can rule out, with high certainty, a red giant companion star" (Kerzendorf et al. 2014), and ``From all the preceding evidence, we can exclude MS, subgiants, giants, and to a certain extent stars below the solar luminosity." (Ruiz-Laputente et al. 2018).  There has been a false alarm when Sun \& Chen (2019) speculated that one particular abundance ratio ([Mg]/[O]) might be produced if the progenitor had a giant companion.  This interpretation is substantially weakened by the authors also proposing a second equally good alternative scenario, ``The abundance ratios from the shocked ejecta are well compatible with the predicted results from spherical delayed-detonation models for Type Ia supernovae."  And the impact of the giant-progenitor speculation is further lowered when they mention the possibility that the ex-companion is a subdwarf B star, with no giant companion in the scenario.  And a fourth speculation is that the supernova was a core-degenerate event, again with no giant companion.  With four suggested scenarios, the giant-companion possibility, with such weak evidence, cannot have any useful confidence.  This does not really matter anyway, because the $M_R$ $>$ $+$3.4 limit has already killed any possibility for a giant ex-companion.  So, despite the false alarm, the deep searches certainly rule out the possibility of a giant companion star and rule out the SSS-progenitor model for Kepler's supernova.

From the literature, I have 6 additional useful ex-companion searches, as listed in Table 1.  Three of these are for LMC supernova remnants, with the limits from my group at the Louisiana State University.  Two of the limits come from big galactic remnants, for which the remnant of the 1006 supernova is the best known.  The limit for SN 1972E in NGC 5253 is for the nearest SNIa outside our Local Group of galaxies, requiring {\it HST} to be pushed to its limits.

For this first test of the SSS-progenitor model, I have nine deep limits for the direct detection of any companion star after the supernova eruption.  All nine limits exclude the possibility that the ex-companion is a red giant star.  So for 9-out-of-9 normal SNIa events, the SSS-progenitor model is excluded as a possibility.  By itself, this result shows that symbiotic stars do not provide any substantial fraction of the SNIa up in the sky.

\subsection{No Red Giant Companion Stars Visible Before the Supernova}

There are two {\it direct} methods for recognizing the red giant companion stars required by the SSS-progenitor model.  One direct method is to look {\it after} the supernova for a surviving ex-companion, as described in the previous subsection.  The second direct method is to look {\it before} the supernova for any companion star while it is still orbiting the WD.  Since no one knows where a nearby supernova will explode, this second method can only be done by looking deep into archival data.  In practice, adequate pre-eruption imaging of supernova sites has only been possible with {\it HST} pictures taken serendipitously\footnote{As {\it JWST} accumulates deep images of nearby galaxies, sooner or later, a SNIa will appear in the field, and {\it JWST} will achieve a detection or deep limit on the companion.}.  This method has been tremendously successful for detecting the red giant progenitors that produce core collapse supernovae (Smartt 2009).  So far, this method has only been applied to two of the nearest SNIa events.

SN 2011fe, in the bright nearby galaxy M101, was discovered by the Palomar Transient Factory within a day of the start of its eruption, and the importance of the event was immediately recognized (Nugent et al. 2011).  Four hours after this initial report, Li et al. (2011a) had already reported that {\it HST} has no visible progenitor star.  The resultant {\it Nature} paper (Li et al. 2011b) was titled ``Exclusion of a luminous red giant as a companion star to the progenitor of supernova SN 2011fe".  Their formal limit is that $M_V$ $>$ $-$1 for hot stars (including V445 Pup).  For stellar temperatures like for red giants, the limit is $M_V$ $>$ 0.0.  This limit does not reject all giant companions, but it does reject all the red giants that are assumed by the SSS-progenitor model (such as in T CrB and RS Oph).

SN 2014J, in the bright and weird starburst galaxy M82, is the closest SNIa discovered since SN 1972E.  The discovery was serendipity by a professor and four students practicing with a telescope in the suburbs of London (Fossey et al. 2014).  {\it HST} has 19 images covering most of the optical and infrared bands, from 1997 to 2010.  No point source at the location of the supernova was detected (Kelly et al. 2014).  These images can be used to calculate limits on $M_V$ that depend on the stellar temperature.  For red giants, this comes to $M_V$ $>$ $-$1.  Again, this limit does not reject all giant companions, but it does reject all the red giants that are assumed by the SSS-progenitor model (such as in T CrB and RS Oph).

In summary, for the direct method of looking for the companion before the supernova, the results are 2-out-of-2 that the progenitors of normal SNIa have no visible companion star to deep limits, sufficiently deep so as to rule out red giant stars like required by the SSS-progenitor model.

\subsection{No Nebular Hydrogen or Helium Emission Lines}

There are five methods that are {\it indirect} for detecting a red giant companion star to a normal SNIa system.  All of these work by seeking inevitable and unique effects that can be caused only by the presence of a companion star near to the exploding WD.  The first two of these indirect methods is testing for the presence of nebular emission lines from hydrogen and helium.  The method for testing for {\it hydrogen} is separate and independent from the method testing for {\it helium}, with the hydrogen-test being the most sensitive for SSS companions, and with the helium-test working for giants stripped of their hydrogen-rich envelopes (like for helium novae, with V445 Pup as the prototype) as well as for all SSS companions.

In the SSS-progenitor model, the companion red giant is blasted by the nearby explosion, with the ablation and stripping of the stellar atmosphere entraining large amounts of hydrogen and helium into the expanding shell (Marietta, Burrows, \& Fryxell 2000, Pan, Ricker, \& Taam 2012).  In particular, any red giant companion will lose 0.3--0.8 solar masses of gas that is hydrogen-rich and helium-rich.  So every SSS-progenitor must have 0.3--0.8 M$_{\odot}$ of H and He shining brightly with emission lines.  Such gas will be visible mainly during the late nebular stages of the eruption due to their bright Balmer and helium emission lines.  For normal SNIa, the ejecta from the WD has no hydrogen or helium, so the only possible source for H or He in the ejecta is to get it from the companion star.  The test is to look for bright H and He lines, and if these are seen, then the progenitor must have had a companion.  If the supernova does not show any H or He lines, then the progenitor can not have had a red giant companion.  So we have a clear and simple means of testing the SSS-progenitor requirement.

The giants in symbiotic stars have roughly solar composition, and the helium will be nearly 25\% by mass with most of the rest being hydrogen.  These giants have a relatively weak surface gravity, meaning that it is easy to remove gas by both ablation (heating) and stripping (momentum transfer).  Detailed hydrodynamic calculations show that the gas entrained into the ejecta ranges from 0.3 M$_{\odot}$ to 0.8 M$_{\odot}$, depending mainly on separation of the companion (Pan, Ricker, \& Taam 2012).  The larger mass is for the S-type symbiotic stars with accretion by Roche lobe overflow, while the smaller mass is for the symbiotic star with the widest known separation that is with wind accretion.  So the SSS-progenitor model requires a minimum of 0.22 M$_{\odot}$ of entrained hydrogen and 0.07 M$_{\odot}$ of entrained helium.

Historically, SNIa are defined, in part, by their lack of any Balmer line emission.  Nevertheless, astronomers have always kept close watch for any hydrogen, especially at late time (e.g. Cumming et al. 1996).  So it was a surprise for SN 2002ic to display a bright and relatively-narrow H$\alpha$ line throughout (Hamuy et al. 2003).  On the face of it, the hydrogen could only come from a massive companion star, and this would seem to prove that something like the SSS model is correct.  But Livio \& Riess (2003) argue that exactly the opposite is the case.  SN 2002ic shows that large amounts of entrained hydrogen are easily detectable, yet it had not been detected in over a hundred prior SNIa, so the case with a giant companion must be rare.  Indeed, SN 2002ic is now recognized as the first example of a rare class of supernova labeled as `Ia-CSM' (Silverman et al. 2013).  Sharma et al. (2023) measure that 0.02\%--0.2\% of normal SNIa are in the Ia-CSM class.  The circumstellar medium around these rare supernova might  have a source of a giant companion star, or possibly as residual debris around a core-degenerate supernova.  The rarity of the Ia-CSM phenomenon, plus the apparent difference in mechanism from the normal SNIa, makes for this class to be irrelevant for the overall progenitor problem for normal SNIa.  And the differences from that predicted by the SSS-progenitor model (e.g.,  that the Ia-CSM have large masses of hydrogen, usually much larger than the mass of any giant in an SSS) precludes the Ia-CSM from being relevant for the SSS-model.  So the Ia-CSM class is just a false alarm for progenitor questions of normal SNIa.

SN 2017cbv and SN 2020nlb are both well-observed normal SNIa events in nearby galaxies.  Like most supernova spectra, they were first checked for hydrogen lines as the usual first diagnostic of the supernova class.  In these two cases, Sand et al. (2018, 2021) looked closely and quantified the limits on hydrogen and helium line flux, and converting these to limits on the entrained hydrogen and helium.  The lines sought are H$\alpha$ and He {\rm I} 6678~\AA.  For SN 2017cbv, the limit on the entrained hydrogen mass is 1$\times$10$^{-4}$ M$_{\odot}$, while the limit on the entrained helium mass is 5$\times$10$^{-4}$ M$_{\odot}$ (Sand et al. 2021).  For SN 2020nlb, the limit on the entrained hydrogen mass is (0.7--2)$\times$10$^{-3}$ M$_{\odot}$, while the limit on the entrained helium mass is 4$\times$10$^{-3}$ M$_{\odot}$ (Sand et al. 2021).  These two supernova have upper limits that are three orders-of-magnitude smaller than the lower limit required by the SSS model.  That is, these two normal SNIa certainly did not have an SSS progenitor.

SN 2018oh was in the field of {\it Kepler} during its K2 Campaign 16, so a nice light curve was taken with continuous imaging for one month with a 30-minute cadence.  The supernova was detected within hours after the explosion, and Dimitriadis et al. (2019b) claim to see a brightening from the Kasen effect (see Section 2.5).  Spectra showed it to be a normal SNIa with no hydrogen or helium lines.  The limit on the entrained hydrogen mass is 5.4$\times$10$^{-4}$ M$_{\odot}$, and the limit on the entrained helium is 4.7$\times$10$^{-4}$ M$_{\odot}$ (Dimitriadis et al. 2019a).

SN 2021aefx, in the nearby spiral galaxy NGC 1566, was discovered within hours of its eruption, provoking a massive effort from Las Cumbres Observatory (in bands {\it UBgVri}), the {\it Swift} (in the three UVOT ultraviolet bands), and the Australia Telescope Compact Array (in two radio frequencies).  This resulted in no detection of hydrogen or helium lines, no detection in the radio, and a claimed detection of a small Kasen effect (Hosseinzadeh et al. 2022).   These limits are in Table 1.

Tucker et al. (2020) collected limits on hydrogen and helium line fluxes for a fair sample of 107 normal SNIa during the nebular phase of the declining light curves.  All of these supernova showed no hydrogen or helium at any level.  These flux limits were translated into mass limits for the entrained hydrogen and helium.  The mass limits for entrained hydrogen get as low as 0.00003 M$_{\odot}$, with a median of 0.0015 M$_{\odot}$.  The mass limits for entrained helium get as low as 0.00004 M$_{\odot}$, with a median of 0.0020 M$_{\odot}$.   These limits are impressive that normal SNIa do not have even a tiny amount of swept up companion atmosphere.   From the entire sample, 107 have limits on the entrained hydrogen $<$0.22 M$_{\odot}$, and 93 have limits on the entrained helium of $<$0.07 M$_{\odot}$.  So out of all the supernova with adequate tests, all 107 have the possibility of a red giant companion excluded.  That is, 107-out-of-107 normal SNIa did not have a red giant companion and certainly cannot have had a SSS-progenitor.

In all, the utter lack of any entrained hydrogen or helium in any of a total of 111 normal SNIa ejecta, to deep limits, means that the SSS-progenitors are at most unmeasurably-rare, being most consistent with a zero fraction of all normal SNIa.  This result, by itself, proves that symbiotic star systems are not the solution of the progenitor problem for normal Type Ia supernova.  And the fraction of normal SNIa that arise from SSS-progenitors must be small, smaller than the $\frac{1}{111}$ level, i.e., $<$0.9\%.

\subsection{No Radio or X-ray Luminosity From the Ejecta Impacting on a Red Giant Wind}

Two further {\it indirect} methods for testing for a companion star are to detect its stellar wind, either in the {\it radio} or {\it X-ray}.  During the explosion, the ejecta will ram into any stellar wind recently emitted from the companion star, producing radiation that is prominent in radio frequencies and in X-ray energies.  The red giants in symbiotic stars, both D-type and S-type in particular, have a heavy wind, so the SSS-progenitor model makes a strong prediction that the SNIa will be extremely luminous radio sources and X-ray sources, starting around the time of the optical peak.  This prediction can be tested.  And there is no known way around this requirement from the SSS model.  

Symbiotic stars are notorious for their heavy stellar winds.  Seaquist \& Taylor (1990) measured the wind strengths from a large survey of radio brightness for 11 D-type and 15 S-type SSSs.  The S-type systems had a median strength of 0.17$\times$10$^{-6}$ M$_{\odot}$ yr$^{-1}$, and a total range of (0.004-0.59)$\times$10$^{-6}$ M$_{\odot}$ yr$^{-1}$.  The D-type systems had a median strength of 2.1$\times$10$^{-6}$ M$_{\odot}$ yr$^{-1}$, and a total range of (0.37-11)$\times$10$^{-6}$ M$_{\odot}$ yr$^{-1}$.  So we can be confident that the wind loss rate from any SSS-progenitor must be $>$0.004$\times$10$^{-6}$ M$_{\odot}$ yr$^{-1}$, with this being the smallest known wind strength for any symbiotic star.  The typical wind velocity ($v_w$) for red giants varies from $\sim$40 km s$^{-1}$ for K2 {\rm III} to $\sim$20 km s$^{-1}$ for M5 {\rm III} (Wood, M\"{u}ller, \& Harper 2016), so I will take 30 km s$^{-1}$ as a good average for all SSS.  

In the SSS-progenitor model, the ejecta/wind impact must be brightly visible in both the radio and X-ray bands.  Many normal SNIa have been deeply studied in the radio with the best technology of the day, and {\it zero} of them have been detected.  This means that SNIa companions do not have any significant stellar winds.  The radio detection efforts have been collected in the paper by Chomiuk et al (2016).  This paper collected archival measures from the Very Large Array (VLA) before its upgrade, collected their own extensive observations with the upgraded Jansky VLA, and collected many observations from the literature.  Their upper limits on the radio flux are converted to limits on the density in the wind, which scales as $\dot{M}_{\rm wind} / v_w$, so as to get a nearly complete sample of radio constraints on $\dot{M}_{\rm wind}$.  From this sample, I have excluded the SNIa from rare classes with likely different physical mechanisms, including Ia-CSM, the low-luminosity events (Iax, Ca-rich SNe, and SN 2002es-like), and the superluminous supernovae.  With this, the Chomiuk et al. sample has 63 normal SNIa.  All 63 have radio limits of $\dot{M}_{\rm wind}$$<$7.5$\times$10$^{-7}$ M$_{\odot}$ yr$^{-1}$ for a 30 km s$^{-1}$ wind, and this is adequate to reject the majority of the uncommon D-type symbiotic stars.  Six of the radio limits are sufficiently deep so as to make a perfect test of the SSS-progenitor model that demands that all such supernova have $\dot{M}_{\rm wind}$$>$4$\times$10$^{-9}$ M$_{\odot}$ yr$^{-1}$.  These deep non-detection limits are presented in Table 1 for six normal SNIa.  The implication of all these radio searches is that normal SNIa are not radio sources to deep limits, so the ejecta/wind impact case has never been seen, so the SSS-progenitor model can at most be rare.

Limits on the presence of any stellar wind can also be gotten from X-ray observations.  Sand et al. (2021) observed SN 2017cbv and SN 2020nlb with {\it Chandra}, in both cases looking deep, as only {\it Chandra} can, and not seeing either supernova.  They have calculated the upper limits for the density of the wind, which scales as $\dot{M}_{\rm wind} / v_w$, and they report these limits for an adopted velocity of 100 km s$^{-1}$.  The appropriate wind velocity for SSS-progenitors is 20--40 km s$^{-1}$, so I have adjusted their published limits to 30 km s$^{-1}$.  The upper limits for the stellar wind are $<$2.2$\times$10$^{-9}$ M$_{\odot}$ yr$^{-1}$ for SN 2017cbv, and $<$2.9$\times$10$^{-9}$ M$_{\odot}$ yr$^{-1}$ for SN 2020nlb (see Table 1).  Both of these {\it upper} limits are significantly smaller than the {\it lower} limit allowed by any symbiotic star companion.  That is, the possibility of a SSS-progenitor is refuted for these two normal SNIa.

Russell \& Imler (2012) report on 53 non-detections in the X-ray band as observed with the {\it Swift} XRT instrument.  Swift has a substantially poorer detection threshold as compared to {\it Chandra}, so the limits are much weaker than those in the previous paragraph.  The quoted {\it Swift} limits on $\dot{M}_{\rm wind}$ need to be corrected to the 1-sigma standard, and need to be corrected to the appropriate wind velocity of 30 km s$^{-1}$.  The median limit is 1$\times$10$^{-5}$ M$_{\odot}$ yr$^{-1}$, with a total range of (0.3--10)$\times$10$^{-5}$ M$_{\odot}$ yr$^{-1}$.  When all 53 supernova are stacked, the joint upper limit is 1$\times$10$^{-6}$ M$_{\odot}$ yr$^{-1}$.  This limit rejects 31\% of symbiotic stars (Seaquist \& Taylor 1990).  I have not included this constraint in Table 1, because its entries are reserved for supernova for which any possibility of a SSS-progenitor can be confidently rejected.

In the end, for the ejecta/wind impact searches, we are left with the stark fact that {\it zero} normal SNIa have ever been detected to have a stellar wind.  This serves as a refutation of all progenitor models that require or use a red giant companion star.  And in particular, these limits demonstrate that SSS-progenitors must be either rare or non-existent.

\subsection{No Kasen Effect of Extreme Brightening in the First Few Days After the Explosion}

The last {\it indirect} method of detecting the companion star is by means of the Kasen effect.  The Kasen effect (Kasen 2010) is when the blastwave from the WD slams into the companion, with the shadowcone of the star creating a hole in the blastwave, with the shocks from the impacting ejecta heating the gas to high luminosity.  The supernova light curve has a sudden rise in brightness starting in the hours after the eruption, continuing for a few days until the hole closes up and the shock-heated gas cools, whereupon the brightness fades back to the usual expanding photosphere light curve.  Radiative diffusion from the inner layers of the shock-heated ejecta makes a separate light curve peak in optical light, lasting for a few days and reaching brightnesses comparable to the main peak around two weeks later.  This blatant `Kasen peak' is prominent for observers looking down the shadowcone, which is for roughly 10\% of the supernovae (Kasen 2010).  The Kasen effect is maximized for red giant companion stars, so this is perfect for testing for SSS-progenitors.  So for roughly 10\% of systems with a red giant companion, we will see a sudden brightening starting around two hours after the explosion with a brightening to absolute magnitudes of $-$17 to $-$18 mag (see Kasen 2010, Figure 3).  With this, we have a simple method of testing for a red giant companion near the line of sight.  Out of 100 supernovae with SSS-progenitors, we should see $\sim$10 with the distinctive immediate rise to near the main-peak luminosity.  If we look at the first day or two of the light curves for 100 normal SNIa and we see the blatant Kasen peaks for 10 supernovae, then this would point to nearly all of the progenitors having a red giant companion.  If we look at the first day or two and see zero Kasen peaks, then this is consistent with the fraction of progenitors with red giant companions being small.  So now we have a good method for testing for red giant companions.

With vast and sustained effort, our community has many effective programs for getting high-cadence images of many galaxies, so that continuous light curves will cover the first hours and days of many SNIa.  By taking frequent images that are both wide-field and deep, many galaxies are covered, and occasional supernovae pop off, with nice coverage from long before the explosion to long into the tail of the light curve.  With efficient analysis pipelines, supernovae are speedily discovered, whereupon the brighter eruptions receive high cadence coverage with many optical and infrared bands, plus coverage in the radio and X-ray for the nearest.  An example of one of these programs is the use of the {\it TESS} full-frame data that continuously take images over 24$\degr$$\times$96$\degr$ fields for $\sim$27 days with  10-or-30 minute time resolution (Fausnaugh et al. 2023).  This {\it TESS} search has light curves for 307 SNIa, with each showing a flat pre-explosion baseline and a clear supernova rise, such that any Kasen peak would be prominent.

The Kasen effect has been reported for six normal SNIa.  None of these are the prominent Kasen peaks where the light curve gets to absolute magnitudes of $-$17 and $-$18 in the first few hours.  Rather, all are very weak signals, inspiring no confidence.  For SN 2018oh, the claimed Kasen effect is not visible in the K2 light curve, but rather appears only as a small bump in the plot of residuals from a power law rise, rising only to 3\% of the main peak (Dmitriadis et al. 2019a, Figure 2).  For SN 2021aefx, the claimed effect is only visible as an apparent inflection in the first day (Hosseinzadeh et al. 2022, Figure 1).  For SN 2023bee, the claimed Kasen effect is not visible in the light curve, rather it is only visible in the plot of residuals from an optimized power-law model (Wang et al. 2024, Figure 4).  The three claimed Kasen effects with {\it TESS} are not even visible in the residual plots, with the effect at maximum only one-tenth of the scatter in the noise (Fausnaugh et al. 2023, Figure 21).  Even if these faint bumps are real, they certainly are not the expected Kasen peaks that should rival the main peak starting suddenly in the first hour after the explosion.

The tiny Kasen effects have several alternative explanations, where no red giant star is involved.  One alternative is that the non-power-law rise in the light curve could simply result from an ordinary excess of $^{56}$Ni in the outer shell of the ejecta (Magee \& Maguire 2020). Another possibility is that the early excess could arise from the ejecta hitting thin circumstellar material, with such close-in debris arising from a wide variety of models, without invoking any red giant companion (Piro \& Morozova 2016).  Another possibility is that the initial ignition of the thick helium shell on the surface of a sub-Chandrasekhar WD will create radioactive material and luminosity in the outer layers of the ejecta (Polin et al. 2019).  With many reasonable alternative, any small Kasen effect cannot be connected to a red giant companion with any useable confidence.

The three supernova with the largest Kasen effect have already been proven to not have any red giant companion stars.  For SN 2001oh, SN 2021aefx, and SN 2023bee, the spectral limits on entrained H and He demonstrate that the companion cannot be a red giant star (Dmitriadis et al. 2019b, Wang et al. 2024, and Hosseinzadeh et al. 2022).  For SN 2021aefx and SN 2023bee, the radio limits demonstrate that the companion cannot be a red giant star (Wang et al. 2024, and Hosseinzadeh et al. 2022).  So the existence of the claimed Kasen effects must arise from some other mechanism, and thus be irrelevant for questions about any SSS-progenitors.

The reported Kasen effects have dubious significance.  There are a variety of data and analysis problems:  First, the existence of the inflections in the light curve can be created or destroyed by varying the power law exponent and the time of the explosion for the model of the baseline light curve, see Figure 4 of Wang et al. (2024).  Second, Fausnaugh et al. (2023) points out that a simple change in the statistical test can verify or deny the inflections.  Third, the best-fit for the Kasen effect on SN 2023bee has its {\it reduced} chi-square of 205.5 (Wang et al. 2024, Table 2), so either the model is terrible or the real photometric error bars are 14$\times$ larger than reported and hence making the purported Kasen effect as insignificant.  Fourth, the three {\it TESS} ``detections are not robust", ``show a slight preference for the addition of a companion interaction to a curve power-law model", and are only claimed as ``tentative interaction candidates" (Fausnaugh et al. 2023).  Fifth, ``correlated noise in the {\it TESS} light curves makes the rate of companion star false positives high" (Fausnaugh et al. 2023).  Sixth, the claimed Kasen effect for SN 2023bee comes with half of the purported Kasen `peak' light curve rejected as being ``significantly compromised" due to intermittent ``bad measurements" from added flux of scattered light (Wang et al. 2024, Figure 11).  With this, there is good reason to think that the claimed inflection in the light curve is just an artifact of the fading tail of the bad data.

So we have many strong reasons for knowing that the reported small Kasen effects have no relevance for questions involving red giant companion stars.  These reasons include that the observed effects are minuscule and completely different from the predicted Kasen peaks being comparable to the main supernova peak.  Also, three good alternative models are waiting to explain any small Kasen effect, so we can have no connection to red giants past a speculative level.  Also, the lack of radio flux, hydrogen lines, and helium lines already proves that the SNIa events claimed to have Kasen effects do not have any red giant companions.  Also, the existence and significance of the claimed Kasen effects are dubious.  So the claimed Kasen effects are all false alarms for any question concerning SSS-progenitors.

Largely on the basis of seeking companion stars, many extensive and exhaustive programs have been run to collect high-cadence light curves including the first hours and days of SNIa explosions.  A product of this is that hundreds of normal SNIa have been tested for Kasen peaks.  For $\sim$10\% of the supernovae (those with our Earth inside the companion's shadowcone), the Kasen peak must be a blatant peak, rising within an hour of the explosion up to absolute magnitudes of $-$17 or $-$18, and lasting for a few days.  A Kasen peak viewed inside the shadowcone will be unmistakable with even the poorest light curve that has coverage back to a few days after the explosion.

Results of searches for Kasen effects are usually reported as summaries for each program.  Fausnaugh et al. (2023) summarized the search of 307 normal SNIa with {\it TESS} with high-cadence light curves of $\sim$27 day coverage from many days before to many days after the explosion.  Zero significant or substantial Kasen effects were seen.  Yao et al. (2019) reports on 122 normal SNIa observed with the Zwicky Transient Factory (ZTF), all with high-quality light curves extending back to near the explosion, all with no glimmer of any Kasen effect.  Hayden et al. (2010) report on their search of 108 confirmed SNIa with well-observed early-time light curves with the Sloan Digital Sky Survey SDSS--{\rm II} Supernova Survey, with zero Kasen effects seen.  Bianco et al. (2011) used top-quality $UBV$ light curves from 87 normal SNIa, as observed with the Supernova Legacy Survey (SNLS), with their lack of Kasen effect made obvious in their Figure 1 that overplots all their data.  Ganeshalingam, Li, \& Filippenko (2011) report on 61 high-quality light curves of normal SNIa as observed by the Lick Observatory Supernova Search (LOSS) program, all with no detectable Kasen effect.  Olling et al. (2015) and Wang et al. (2021) report on 4 normal SNIa observed with a continuous 30-minute cadence from the {\it Kepler} spacecraft, with no significant excess flux at early times.

In all, 689 normal SNIa have published searches for Kasen peaks, and zero are seen.  With only $\sim$10\% of SSS-progenitors expected to display the blatant peaks, the SSS model requires $\sim$69 to have Kasen peaks, if all normal SNIa have SSS-progenitors.  Still, we can consider that some small fraction of the normal SNIa arise from symbiotic stars, and the statistics on the Kasen effect can limit the fraction.  For example, if 10\% of the normal SNIa have symbiotic progenitors, then $\sim$7 out of the entire sample of 689 events should show blatant Kasen peaks, for which the Poisson probability is 0.0010 that zero will be seen, so the case is ruled out at more than the 3-$\sigma$ confidence level.  The Poisson probability of getting zero detections better than the 1-$\sigma$ level only when the SSS fraction is at the $<$$\frac{1}{69}$ level.  So with just the Kasen effect result, the SSS-progenitor fraction amongst normal SNIa is $<$1.4\%.

\subsection{Conclusions on Red Giant Companion Searches}

The stark facts remain that {\bf (A)} zero normal SNIa have ever been seen to have any ex-companion near the center of a supernova remnant, {\bf (B)} zero normal SNIa have ever been seen as a pre-eruption progenitor, {\bf (C)} zero normal SNIa have ever been seen to have hydrogen lines in their eruption spectra, {\bf (D)} zero normal SNIa have ever been seen to have helium in their eruption spectra, {\bf (E)} zero normal SNIa have ever been detected in the radio, {\bf (F)} zero normal SNIa have ever been detected in the X-rays, and {\bf (G)} zero normal SNIa have ever had any substantial brightness peak in the first few days.  The answer is obvious.  SNIa do not have red giant companion stars in any measurable fraction.  Symbiotic stars do not contribute any measurable fraction of Type Ia supernova progenitors.

We can be quantitative about the limits on the fraction of SSS-progenitors for normal SNIa.  In Table 1, the third column has a substantial number of duplicates, for example, the 93 systems with nebular He limits from Tucker are all duplicates of the 107 systems with nebular hydrogen lines from Tucker.  For the lines above the Kasen effect entries, there are a total of 120 individual SNIa for which the possibility of a SSS-progenitor has been confidently ruled out.  A total of 689 systems have had published checks for the Kasen effect on top-quality light curves with good early-time coverage.  Of the 689 systems with published test results, something like 10\% will have the Earth inside the shadowcone, so the tests will be sensitive to red giant companions.  So $\sim$69 supernova have been tested for the Kasen effect of a red giant companion.  This brings a total of around 189 supernovae that have been tested for a SSS-progenitor.

Out of the 189 tested normal SNIa, zero have a red giant companion.  All the supernovae tested in Table 1 have highly confident test results that no red giant companion is present.  There is no possibility that any one of the 189 had a giant star in the progenitor.

With zero-out-of-189, we can get a quantitative limit on the fraction of normal SNIa that have SSS-progenitors.  The best estimate is for a SSS-fraction of 0.0\%.  If the fraction exactly equals zero, then the SSS-model would be impossible.  However, it is formally possible that the SSS-fraction is simply very small while being non-zero, which is to suggest that SSS-progenitors are rare, but not impossible.  Quantitatively, the SSS-progenitor fraction is formally $<$$\frac{1}{189}$, or $<$0.53\%.

\section{CONCLUSIONS}

If the immediate SNIa progenitor is a symbiotic star, then the companion star must be a red giant with a heavy stellar wind.  The contrapositive is the identical requirement that if the companion is not a red giant with a massive stellar wind, then the progenitor is not a symbiotic star.  So the test is to look at normal SNIa to see whether the companion is a red giant or has a heavy stellar wind.  There are seven methods which can be used to test the cases.  For all seven methods (A--G below), a total of 189 different normal Type Ia supernovae have been examined with the all-189 being proven to not not have any red giant companion with a strong stellar wind.  This is proof from many widely-different techniques that the fraction of normal SNIa with SSS-progenitors is $<$0.53\%.

{\bf (A)} All red giant companion stars can be directly detected by looking near the center of the supernova remnant.  Such ex-companion stars will still have nearly their original luminosity and surface temperature.  For this test, we can only use remnants that are provably from a normal SNIa (from their eruption or light echo spectra), that are recent enough (within roughly 5 centuries) so that the red giant has not had time to move far from the explosion site, and must be near enough to recognize the ex-companion.  These conditions are present for only four galactic supernova remnants, four remnants in the LMC, and in the one closest normal SNIa to our Local Group.  For all 9 remnants, no red giant is ever found anywhere near the center.  This test is definitive for each of the nine remnants, because there is no way to hide the ex-companion.  With zero-out-of-9 testable remnants possibly having a red giant ex-companion, we know that these nine never had a SSS-progenitor.  Already, this one result proves that the SSS-progenitor rate is $<$11\%.  This one result is adequate for our community to know that symbiotic stars do not provide normal SNIa progenitors at a rate so as to solve the progenitor problem.

{\bf (B)} A second direct method for detecting the red giant companion is to look at archival pre-explosion images.  In practice, this test is possible only for the nearest normal SNIa, with {\it HST} providing the deep archival images.  As recent and very nearby supernovae, these events have excellent spectra to prove the normal-SNIa classification, and positions to 21 milli-arc-second accuracy.  For the two testable cases (SN 2011fe and SN 2014J), the limits on the progenitor brightness are $M_V$ less luminous than 0.0 and $-$1, respectively.  These limits rule out any possibility of a red giant companion, and so rule out any SSS-progenitor for these two systems.  This result has importance because it is a second independent direct test of red giants, and one for which there there is no realistic way to avoid the conclusion.

{\bf (C)} An indirect method of testing for a red giant is to look for the inevitable hydrogen gas stripped off the companion by the supernova blastwave and then entrained into the ejecta to emit strong Balmer lines.  Bright H$\alpha$ emission is an inevitable consequence of the SSS-progenitor model, where $>$0.22 M$_{\odot}$ must get entrained.  Bright H$\alpha$ lines should be seen in all supernovae that had SSS-progenitors, yet such has been seen in zero normal\footnote{Normal SNIa do not include the types Iax, Ia-CSM, or the superluminous supernova, for which the Ia-CSM do show Balmer lines.  But these classes apparently involve some completely different eruption scenario and mechanism, thus making them irrelevant to the SNIa progenitor problem.  Further, these classes are rare, with these being $<$0.2\% of the rate of normal SNIa (Sharma et al. 2023), so such being irrelevant to the SNIa progenitor problem.} SNIa.  For the best observed cases, the limit on stripped hydrogen mass is $<$0.000033 M$_{\odot}$, nearly four orders of magnitude below the minimum required by the SSS model.  For published analyses that convert flux limits on the Balmer lines to limits on the entrained hydrogen, all 111 cases show no Balmer flux.  By itself, this result is a confident proof that $<$0.9\% of normal SNIa have SSS-progenitors.  The real limit is greatly tighter than this, because thousands of normal SNIa have been examined for H$\alpha$ (as part of the normal classification process), even modest spectra would have revealed the line flux with limits deeper than 0.22 M$_{\odot}$, and any such detection would have been speedily published.  So the real number of normal SNIa with no hydrogen lines is $\gg$1000, and the fraction of SSS-progenitors is $\ll$0.1\%.  The utter lack of hydrogen lines in any of hundreds or thousands of normal SNIa is proof that the symbiotic stars do not provide any measurable fraction of the normal supernovae.

{\bf (D)} A related indirect method for red-giant-testing is to look for the inevitable helium gas stripped off the companion by the blastwave, entrained into the ejecta, and shining brightly as a helium emission line.  Any progenitor from the SSS scenario must entrain $>$0.07 M$_{\odot}$ of helium, and this will shine brightly with {\rm He I} emission lines, mainly at 5875~\AA~ and 6678~\AA.  So we have a simple test for SSS-progenitors, which can be made for most supernovae, and for which I know of no way to confound the test.  For this test, the best observed normal SNIa have limits on entrained helium of $<$0.000040 M$_{\odot}$, over four orders-of-magnitude smaller than required by the SSS-model.  For the published limits on entrained helium, all 97 have no detection of helium to limits of 0.07 M$_{\odot}$ or deeper.  This test is similar as for the entrained-hydrogen limits, yet it comprises an independent measure to provide yet more confidence in the basic result.  Further, the entrained-helium test covers any SD models involving helium giants, for example helium novae\footnote{The only known case of a helium nova is V445 Pup, so numerous papers have put forth this system as the exemplar of the helium-nova progenitor.  Schaefer (2025) has been able to prove by multiple methods that V445 Pup is not a SNIa progenitor.  Now, the limits on entrained-helium proves that the helium nova channel does not contribute any measurable fraction to the observed normal SNIa.}.  The utter lack of any helium emission lines in the spectra of hundreds of normal SNIa proves that the SSS progenitors are not making any of the SNIa that we see up in the sky.

{\bf (E)} A strong indirect method for testing the SSS model is to seek evidence for a strong stellar wind in the progenitor.  Amongst progenitor models, the SSS scenario is unique for having/requiring a dense wind, and all SSS winds have strengths $>$4$\times$10$^{-9}$ M$_{\odot}$ yr$^{-1}$, and usually with wind strengths that are orders-of-magnitude larger.  This strong wind will necessarily produce a highly luminous radio source arising from the impact of the supernova ejecta with the wind around the red giant companion.  So the test of the SSS model is to seek a bright radio source from normal SNIa in the weeks and months after the explosion.  To date, no normal SNIa has ever been detected as a radio source, despite deep searches of many events.  As of 2014, a total of 6 normal SNIa have published limits that reject even the extreme SSS case.   I am not aware of any way to impeach this method involving radio detection, so this method provides a confident rejection of the presence of any stellar wind around the companion.  Therefore, the lack of radio detections demonstrates yet another greatly different technique that proves that normal SNIa do not arise from SSS in any significant fraction.

{\bf (F)} A similar indirect method to test the SSS model is to seek the inevitable X-rays from the ejecta/wind impact that must arise in all SSS-progenitors.  That  is, in the SSS model, the red giant must have a wind of strength $>$4$\times$10$^{-9}$ M$_{\odot}$ yr$^{-1}$, and such must produce a strong X-ray source in the days and months after the explosion.  However, no normal SNIa has ever been detected in the X-rays.  I have been able to find only two cases where the X-ray flux limit has been turned into a limit on the wind strength in the progenitor.  In both cases, the observed limit is $<$2.9$\times$10$^{-9}$ M$_{\odot}$ yr$^{-1}$, which eliminates any possibility for a SSS progenitor.  While the numbers for this method are not large, the X-ray method does provide yet another confirmation by a greatly different set of observers, instruments, physics of emission, and calculation of limits.  Yet again, we have a conclusion that SNIa progenitors do not have a red giant with a strong wind, so the SSS are not progenitors.

{\bf (G)} A completely different indirect method to test for symbiotic progenitors is to seek the Kasen effect, wherein roughly ten-percent of the progenitors with a red giant companion star (those viewed down the shadowcone) must have a fast rise over the first few hours after the explosion with a brightening to near the main peak level.  Such a blatant rise would be easily visible in any light curve that covers the first few days.  But no such Kasen effect has ever been seen.  I tabulate 689 supernova that have published searches for the Kasen effect, for which zero have any significant rise in the first few hours.  Out of these 689 SNIa, $\sim$10\% will have the blatant Kasen peak, or $\sim$69, if all have SSS-progenitors.  That zero are seen is proof that the SSS model is not the solution to the progenitor problem.  This is just one of seven greatly different methods, where every time that astronomers have tested a normal SNIa for the presence of a red giant companion with a strong wind, the SSS-model fails.

In summary, all normal SNIa show no evidence of the required red giant companion with a strong wind.  With this, normal SNIa do not come from SSS-progenitors.  In conclusion, the SSS model does not actually provide any measurable fraction of normal SNIa.

\begin{acknowledgments}
I thank Scott Kenyon (Center for Astrophysics) for comments and advise on the science in this paper.  I appreciate the comments and help from Joanna Miko{\l}ajewska (Nicolaus Copernicus Astronomical Centre) on the science content.
\end{acknowledgments}

%



{}


\end{document}